# High quality factor nanophotonic resonators in bulk rare-earth doped crystals

TIAN ZHONG,[1,2] JAKE ROCHMAN,[1] JONATHAN M. KINDEM,[1,2] EVAN MIYAZONO,[1,2] ANDREI FARAON.[1,2*]

[1]*T. J. Watson Laboratory of Applied Physics and Kavli Nanoscience Institute, California Institute of Technology, 1200 E California Blvd, Pasadena, CA, 91125, USA*
[2] *Institute for Quantum Information and Matter, California Institute of Technology, Pasadena, California 91125, USA*

*Corresponding author: faraon@caltech.edu*

**Numerous bulk crystalline materials exhibit attractive nonlinear and luminescent properties for classical and quantum optical applications. A chip-scale platform for high quality factor optical nanocavities in these materials will enable new optoelectronic devices and quantum light-matter interfaces. In this article, photonic crystal nanobeam resonators fabricated using focused ion beam milling in bulk insulators, such as rare-earth doped yttrium orthosilicate and yttrium vanadate, are demonstrated. Operation in the visible, near infrared, and telecom wavelengths with quality factors up to 27,000 and optical mode volumes close to one cubic wavelength is measured. These devices enable new nanolasers, on-chip quantum optical memories, single photon sources, and non-linear devices at low photon numbers based on rare-earth ions. The techniques are also applicable to other luminescent centers and crystals.**

## 1. Introduction

Optical nanocavities with high quality factors and small mode volumes are an enabling technology for on-chip photonic devices such as low-power opto-electronic switches, low threshold lasers, cavity-optomechanics, and on-chip quantum information processing [1-3]. In particular, nanoresonators with a large quality factor-to-mode volume ratio are desirable for strong Purcell enhancement of light-matter interactions that leads to high optical nonlinearity [4], efficient lasing [5,6], bright quantum light emissions [7, 8], and opto-electronic devices operating at the single photon level [9,10]. Most nanophotonic platforms are based on commonly used semiconductor materials such as silicon, gallium arsenide and indium phosphide. However, there are numerous other materials, including complex oxide crystals (e. g. yttrium orthosilicate (YSO), yttrium vanadate (YVO), lithium niobate ($LiNbO_3$), and potassium titanyl phosphate ($KTiOPO_4$)), with interesting nonlinear and luminescent properties that can be exploited for applications in classical and quantum nano-optics. Nano-fabrication techniques for these materials are very limited due to the unavailability of either selective etching chemistries or high-quality thin films on which photonic devices can be made. Several attempts to fabricate nanocavities on unconventional materials, such as single-crystal diamond [8, 11] or lithium niobate crystals [12], have been successful, but their methods cannot be easily transferred to other substrates. On the other hand, focused ion beam (FIB) milling provides a universal tool for micromachining virtually any bulk materials without requiring thin films. While several studies on fabricating photonic cavities using FIB have been carried out, the results have mostly been of low quality factors [13, 14, 15], which have been attributed to optical property degradation, unrepeatable patterning or significant material stress. Specifically, Ref. [13] demonstrated a nanobeam photonic crystal cavity in diamond with a maximum Q of 221. Ref. [14] employed a diamond thin film on which both 1-D and 2-D photonic crystal cavities were milled. The achieved Q ranged from 180 to 700. Lastly, Ref. [15] obtained a Q of 900 by milling a point defect in a silicon-on-insulator slab cavity.

In this work we demonstrate a new design and a robust fabrication platform for nanophotonic resonators based on triangular nanobeams with longitudinal grooves milled in bulk complex oxide crystals (YSO and YVO), which are common hosts for rare-earth emitters. Unlike a common photonic crystal design based on air holes in a thin slab of substrate, the triangular nanobeam geometry combined with a rectangular subwavelength groove optical lattice results in better tolerances to drifts perpendicular to the nano-beam direction that occur during the fabrication process. The fabricated devices exhibit high quality factors up to 27,000 and small mode volumes of ~1 $(\lambda/n)^3$ over a wide spectrum range, from visible to near infrared, to telecom, with resonance wavelengths closely matched to atomic transitions of multiple rare-earth ion dopants including Europium (Eu), Praseodymium (Pr), Neodymium (Nd) and Erbium (Er). The quality factors of these devices are several times better than the highest Qs reported to date in FIB-machined cavities [14].

## 2. Design

Many photonic crystal cavity designs use circular perforations in the center of a 1-D nanobeam to create a photonic bandgap and use modulations of the perforation's geometry to obtain a local cavity mode within the structure [16-18]. When using focused ion beam milling, this method presents difficulty for making high-Q cavities largely due to misalignment of the perforations with respect to the nanobeam axis. To minimize this issue, a photonic crystal cavity design consisting of a triangular nanobeam [13] with a lattice of sub-wavelength grooves, shown in Fig. 1, is proposed. The pattern of grooves on the triangular beam is milled across the entire beam width, which eases the alignment of the grooves to the nanobeam and permits significant design robustness against fabrication errors due to ion beam drift during milling. Another benefit of the triangular nanobeam design is that all its dimensions can be globally scaled to attain resonances in a wide spectral range on the same chip. This flexibility is generally less available for photonic devices fabricated on thin films, for which a different thin film thickness is required to match specific resonance wavelengths.

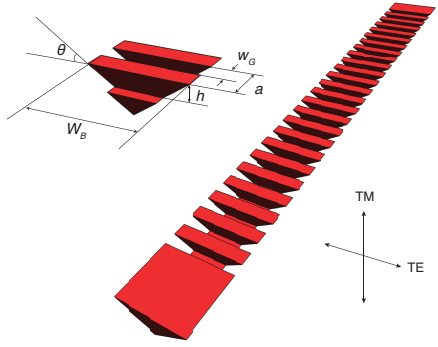

Fig. 1. Schematic of the triangular nanobeam resonator. The zoom-in view shows the photonic crystal structure of the optical lattices labeled with relevant design parameters.

As seen in Fig. 1, the proposed photonic crystal structure can be characterized by the following parameters: beam width $w_B$, lattice constant $a$, groove width $w_G$, groove depth $h$, and nanobeam interior angle $\theta$. The cavity defect was implemented by perturbing the lattice spacing to reach a value of $0.95a$ in the center of the cavity. To minimize scattering and maintain a small mode volume, the lattice spacing was quadratically modulated over 7 lattice spacings from the nominal value $a$ to $0.95a$ at the center of the cavity. A total of 40 grooves are milled into the beam (14 of which have a perturbed lattice spacing), which permits a high-Q design while maintaining reasonable collection and excitation of light through the cavity mirrors. A beam interior angle of 60º is implemented to maximize the symmetry of the mode profile for ease of coupling to the cavity mode through the cavity mirrors. The groove depth $h$ was 70% of the total beam depth to obtain a photonic band gap while maintaining the mechanical integrity of the beam.

## 3. Methods

### A. Device Fabrication

YSO (Scientific Materials Corp.) and YVO (United Crystals) crystals were coated with a 50 nm chrome film using an electron beam evaporator to provide a charge-dissipation and hard mask layer for subsequent milling. The nanobeam cavities were milled in their respective crystals with a focused Ga+ ion beam (FEI NOVA 600). First, a long beam with a triangular cross-section was released from the substrate by milling two rectangular features using a high current at an incidence angle of 30º. The triangular nanobeam was thinned down to the target beam width by gradually milling material (with a low current) on both sides of the beam to minimize the sidewall roughness and remove re-deposited material on the sidewalls. Once the target beam width was obtained, the sample was rotated such that the ion beam was normal to the crystal for patterning of the photonic crystal grooves, with a low current. Rectangular grooves were milled across the entire nanobeam, where the groove width, depth, and lattice spacing contributed the photonic crystal character. All nanobeam milling was monitored in-situ via SEM imaging. Two coupling ports were milled at both ends of the beam at 45º with respect to the sample surface to permit broadband excitation and collection through the cavity mirrors. All milling was performed with an ion beam voltage of 20 keV and at the lowest possible current while maintaining a reasonable milling time (typically 23-760 pA for milling times under 10 minutes per elementary pattern). For instance, to fabricate a 883 nm YSO resonator, 760 pA milling current was first used for quick release of the triangular beam, followed by a 37 pA to trim the beam width down to its design values. The last step used a 23 pA current to pattern the grooves. The total time for making one such device was ~30 minutes. Finally, the chrome layer was removed with chrome etchant (CR-7S) before optical characterization of the devices.

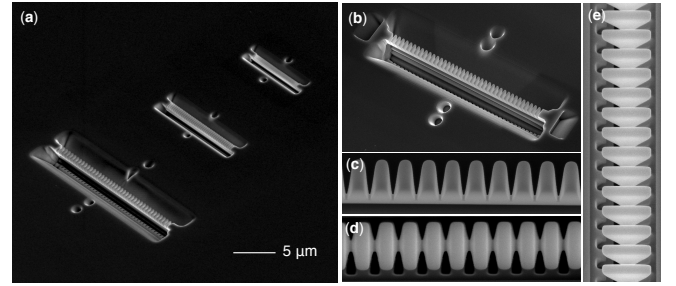

Fig. 2. Scanning electron microscope images of the fabricated nanobeam resonators. (a) Devices for different spectrum range with identical structure features but different global scaling factors. (b) The fabricated device in YVO crystal, which has the same geometric structures as devices in YSO. The side-view in (c) shows the non-vertical sidewalls due to FIB beam divergence, which can be improved by varying the ion beam voltage and current. The top-view in (d) reveals the grooves on a thin support beam. The triangular cross-section of the beam is seen in (e).

Figure 2 (a) shows a SEM image of three fabricated nanobeam resonators for operation at visible, near infrared and telecom wavelengths. All three devices share the identical design and differ only by a global scaling factor. A typical nanobeam resonator with two coupling ports can be seen in Fig. 2 (b). Here, the target resonance wavelength is 1064 nm, but the structure is representative of all fabricated cavities. The high-aspect ratio grooves with an estimated sidewall angle of 6º are shown in Fig 2 (c). This angle is largely characteristic of the focused ion beam parameters implemented, e.g. current, voltage, and beam alignment. Grooves that extend across the entire beam width, and a thin support beam with a width of $0.3w_B$ underneath the grooves can be observed from the top view of the cavity in Fig. 2 (d). The triangular shape of the beam, with an interior angle of 60º, can be seen in Fig. 2 (e).

### B. Optical Measurement

Cavity transmission spectra were measured with a custom-built microscope, where the input from a supercontinuum source was focused with a 50x microscope objective (NA=0.65) onto the 45º-angled coupler at the end of the beam. The coupler reflects the input light propagating normal to the crystal surface into the waveguide. The couplers rely on total internal reflection and have minimal dependence on wavelength, which allows broadband coupling to the nanobeams. The coupler efficiency was measured to be ~20% from transmission measurement in a bare nanobeam without grooves. Transmitted light was collected from the other coupler at the opposite end of the nanobeam and passed through a pinhole at the output path to spatially filter the transmitted light for measurement with a spectrometer. We also measured a 25% coupling efficiency of the output light from the cavity into a single-mode fiber.

## 4. Results

### A. YSO Photonic Crystal Nanobeam Cavities

YSO triangular photonic crystal nanobeams were simulated with MEEP [19], a finite-difference time-domain (FDTD) solver, to optimize the quality factor (Q) by sweeping the values for $w_B$, $a$, $w_G$, and $h$. The simulations used actual geometries of the milled structure from SEM measurements, which included the 6º-sloped sidewalls of the grooves. A refractive index of $n_{YSO}$=1.8 was used for the YSO TE-polarized resonance mode. The optimal design parameters were a beam width of $w_B$ = 0.93 λ, a groove width of $w_G$ = 0.227 λ, and a lattice spacing of $a$ = 0.386 λ. The resultant theoretical Q-factor is ~$7 \times 10^4$ with a mode volume V ~1.6 ($\lambda/n_{YSO}$)$^3$ for the fundamental resonance mode as shown in Fig. 3. The mode is largely confined within the defect region, where the top-view, side-view, and cross-section field profiles are shown in Fig. 3 (a-c), respectively. Since the structure's dimensions scale globally, the same design was used for cavities matching the visible $^3H_4$-$^1D_2$ transition in Pr$^{3+}$:YSO (605 nm), the near-infrared $^4I_{9/2}$-$^4F_{3/2}$ transition in Nd$^{3+}$:YSO (883 nm), and the telecom $^4I_{15/2}$-$^4I_{13/2}$ transition in Er$^{3+}$:YSO (1536 nm) [20].

a lattice spacing of $a$ = 0.352 $\lambda$, which resulted in a theoretical Q-factor of ~3×10$^5$ and mode volume V~1($\lambda/n_{YVO}$)$^3$. A significantly higher quality factor is predicted here due to the larger refractive index. A +/- 5% change in any geometric parameter of the nanocavity had minimal effect on the simulated quality factors, which remained above 10$^5$. This optimized set of parameters was used to fabricate cavities with target wavelengths of 880 nm for the $^4F_{3/2}$-$^4I_{9/2}$ quantum memory transition [21], and 1064 nm for the $^4F_{3/2}$-$^4I_{11/2}$ lasing transitions in Nd$^{3+}$:YVO by globally scaling all the dimensions.

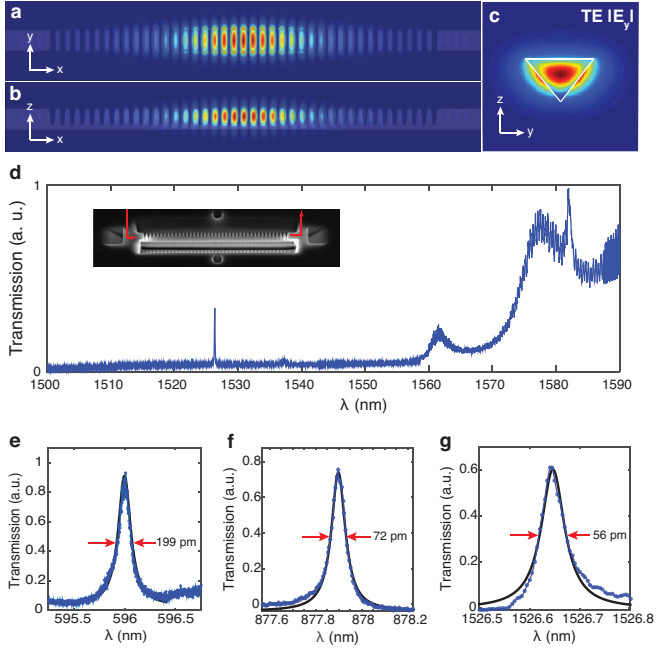

Fig. 3. Nanobeam resonators in the YSO crystal. (a-c) Top, side and cross-section views of the simulated mode profiles of the TE-mode resonance. (d) Typical TE broadband transmission spectrum of a YSO nanobeam resonators showing a resonance in the photonic bandgap. Inset shows the transmission measurement scheme in which broadband super-continuum light vertically couples into the nanobeam from one end and is collected from the other end. (e) Resonance close to the target 605 nm atomic transition of Pr$^{3+}$ ions. (f) Resonance close to the 883 nm transition of Nd$^{3+}$ ions. (g) Resonance close to the 1536 nm transition of Er$^{3+}$ ions.

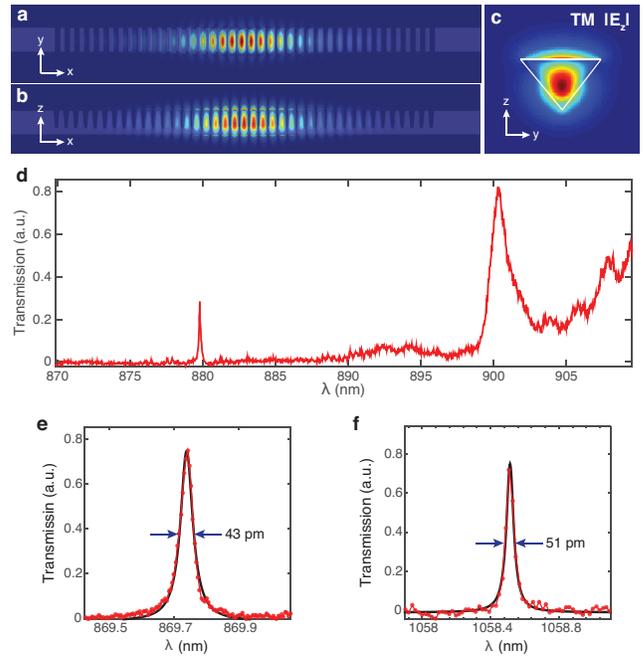

Fig. 4. Nanobeam resonators in the YVO crystal. (a-c) Top, side and cross-section views of the simulated mode profiles of the TM-mode resonance. (d) Typical TM broadband transmission spectrum of YVO nanobeam resonators showing a resonance in the photonic bandgap. (e) Resonance close to the target 880 nm atomic transition of Nd$^{3+}$ ions. (f) Resonance close to the target 1064 nm lasing transition of Nd$^{3+}$ ions.

Transmission spectra of three independent YSO nanocavities for Pr, Nd, and Er transitions are shown in Fig. 3 (e-f). A Q~3000 was measured at $\lambda$ = 596 nm in the device designed for Pr dopant. On a second resonator for operation in near infrared, a higher Q of ~12,000 was measured at $\lambda$ = 877.8 nm. On a third nanobeam scaled up for operation at telecom wavelengths, the highest Q of ~27,000 was measured at $\lambda$ = 1526.6 nm. The increasing Q factors with resonance wavelengths are expected, as scattering loss due to surface roughness in those larger devices is less significant.

**B. YVO Photonic Crystal Nanobeam Cavities**

Nonlinear optical processes often have strong polarization selectivity. Quantum emitters embedded in crystals also have preferred dipole orientations that should match with the cavity polarization. It is thus desirable for the nanocavity to support both TE and TM-polarized modes. Here we investigated YVO triangular nanobeam resonators using a different set of design parameters than the YSO cavities, because of a larger refractive index n$_{YVO}$=2.2 along the c-crystallographic axis of YVO. A TM resonance mode is designed to align to the c-axis, which also coincides with the strongest dipole axis of neodymium ions in the YVO crystal. The optimal design parameters were a beam width of $w_B$ = 0.852 $\lambda$, a groove width of $w_G$ = 0.210 $\lambda$, and

The measured transmission spectra of two YVO cavities are shown in Fig. 4. A Q of ~20,000 at $\lambda$ = 869.7 nm was measured as shown in Fig. 4(e). In a second YVO cavity scaled for the 1064 nm lasing transition, a spectrometer-limited Q of ~20,700 at $\lambda$ = 1058.5 nm was measured.

## 5. Discussion

The design and characterization parameters for all the devices investigated are summarized in Table 1. As important figures of merit, the Q/V (V is in units of ($\lambda$/n)$^3$) ratios are shown in the last column, of which a peak value of 16,870 is obtained for the TE mode in YSO and 20,700 for TM mode in YVO. Following an analysis similar to the one in [20], where materials like Pr$^{3+}$:YSO, Nd$^{3+}$:YSO, Er$^{3+}$:YSO and Nd$^{3+}$:YVO have been theoretically analyzed in detail, the single ion cooperativity is greater than one when Q/V has a value of a few thousands, as already achieved in this work. Using the devices presented in this paper, the singe ion coupling rate g would range from ~1 MHz for Er$^{3+}$ in YSO to ~25 MHz for Nd$^{3+}$ in YVO. The high Q/V values in these resonators fulfill one of the essential requirements for achieving coherent light-matter interactions in the cavity. The other important factor is that the fabricated nano-structures surrounding the ions should preserve the same properties as high quality bulk materials. In that regard, the high Qs of the fabricated devices already suggest that no significant optical property degradation was present. Additionally, we have recently shown in [22] an excellent preservation of the narrow inhomogeneous linewidths and the long optical coherence times (up to 100 μs) of $Nd^{3+}$ ions in one of these YSO nanocavities (similar to no. 2 cavity in Table 1), confirming minimal crystal damage due to ion beam milling. The high quality optical nanocavities demonstrated here combined with the ability to preserve emitters' coherent properties [22] indicate a strong prospect of this platform for scalable quantum information applications, including ensemble-based quantum memories [21] and single rare-earth-ion qubits [20].

Table 1: Summary of parameters for nanobeam resonators fabricated in YSO and YVO crystals.

| Cavity | Host Material (Polarization) | Resonant λ (nm) (atomic transition) | Beam width $W_B$ (nm) | Lattice a (nm) | Groove width $W_G$ (nm) | Q-factor ($Q_{th}$) | V (μm³) | Q/V (λ/n)⁻³ ($Q_{th}/V$) |
|---|---|---|---|---|---|---|---|---|
| 1 | YSO (TE) | 595.8 (605) | 600 | 250 | 145 | 3,000 (70,000) | 0.058 | 1875 (43,750) |
| 2 | YSO (TE) | 877.9 (883) | 820 | 340 | 200 | 12,200 (70,000) | 0.186 | 7,625 (43,750) |
| 3 | YSO (TE) | 1526.6 (1536) | 1530 | 580 | 360 | 27,000 (70,000) | 0.977 | 16,875 (43,750) |
| 4 | YVO (TM) | 869.7 (880) | 740 | 310 | 185 | 20,000 (300,000) | 0.062 | 20,000 (300,000) |
| 5 | YVO (TM) | 1058.5 (1064) | 880 | 370 | 215 | 20,700 (300,000) | 0.112 | 20,700 (300,000) |

Fabricated devices currently have average quality factors ~10 times lower than theoretical predictions ($Q_{th}$). The deviation can be attributed to imprecision in the geometry, materials absorption or surface roughness. Possible methods to increase the quality factors include decreasing the sidewall angle [23, 24] and post-fabrication annealing [25]. For example, a slight change in the sidewall angle from 6° to 4° can increase the theoretical quality factor of the YSO design by a factor of ~2 from 70,000 to 150,000 in YSO. Further optimization of the ion beam parameters could make such an angle possible. Previous reports with other materials suggest that post-fabrication annealing can decrease material absorption [25], which may lead to increased quality factors. Effects of annealing on FIB nanocavities have yet to be studied with current samples, but could be implemented for improved material quality. It should also be noted that here devices were fabricated in materials with a relatively low refractive index. Materials with a higher refractive index are likely to produce nanocavities with higher quality factors using the proposed design.

Lastly, it is noted that the cavity resonance is generally +/- 10 nm of the target wavelength. This demonstrates repeatable fabrication of nanocavities at specified target resonant wavelengths, a requirement for quantum optical applications. Within +/- 10 nm, several techniques can be used to tune the resonance, such as gas tuning for red-shifting or isotropic etching for blue-shifting. We tested the $N_2$ gas tuning technique on the resonators at cryogenic temperatures (4 K) [22], and found that the resonances at ~880 nm can be continuously red-shifted up to 15 nm in YSO devices without noticeable degradation of Q. The $N_2$ gas tuning range for YVO devices operating at 880 nm is ~10 nm.

## 6. Conclusion

In summary, we have demonstrated a new photonic crystal nanocavity design and fabrication scheme to produce devices with high Q factors and small mode volumes. Quality factors approaching 30,000 at telecom and 12,000 at near infrared wavelengths in low-index YSO with a TE-polarized mode were demonstrated. Cavities with a quality factor exceeding 20,000 at infrared wavelengths were measured in YVO with a TM-polarized mode. To the best of our knowledge, these devices provide the highest Q/V ratios using FIB fabrication scheme. The technique demonstrates versatility in terms of material, wavelength and cavity mode polarization, opening the doors for various optical materials to be studied and utilized in a nanophotonic platform.

**Funding**. NSF CAREER 1454607, NSF Institute for Quantum Information and Matter PHY-1125565 with support from Gordon and Betty Moore Foundation GBMF-12500028, AFOSR Young Investigator Award FA9550-15-1-0252, AFOSR Quantum Transduction MURI #FA9550-15-1-002. Device fabrication was performed in the Kavli Nanoscience Institute with support from Gordon and Betty Moore Foundation.

**Acknowledgment**. T. Zhong and J. Rochman contributed equally to this work. J. Rochman is currently at University of Waterloo, Canada.